\documentclass[prl,twocolumn]{revtex4}
\pdfoutput=1
\usepackage{graphicx}

\begin{document} 

\title{A microscopic description of light induced defects in amorphous silicon solar cells}

\author {Lucas K. Wagner and Jeffrey C. Grossman } 
\affiliation { 
Berkeley Nanosciences and Nanoengineering Institute, University of Californa, Berkeley  Berkeley, CA 94720 \\
E-mail:  lkwagner@berkeley.edu, jgrossman@berkeley.edu
}

\date{\today}

\begin{abstract}
Using a combination of quantum and classical computational approaches, we 
model the electronic structure in amorphous silicon in order gain understanding of
the microscopic atomic configurations responsible for light induced degradation
 of solar cells.  We demonstrate that regions of strained silicon bonds could be
as important as dangling bonds for creating traps for charge carriers.  Further, our results 
show that defects 
are preferentially formed when a region in the amorphous silicon contains 
\emph{both} a hole and a light-induced excitation. These results agree with the puzzling dependencies on 
temperature, time, and pressure observed experimentally. 
\end{abstract}


\maketitle

Despite great promise as an inexpensive and efficient solar cell material\cite{thin_film_review},
hydrogenated amorphous silicon(a-Si:H) is severely limited by the
Staebler-Wronski effect (SWE)\cite{swe}, in 
which the efficiency is degraded by 25-30\% within a few hours of exposure to light.
While there is a strong consensus \cite{stutzmann}
that unsaturated (dangling) silicon bonds play an important role in the SWE by acting as charge traps,
it is also clear that they cannot explain the entire effect 
\cite{fritzche_review}. 
In fact, the photoconductivity can vary by more than a factor of ten at the same 
density of dangling bonds\cite{shimizu_review}.

In experiments, it has been observed\cite{hydrogen_weak_bonds,disorder_prl,hydrogen_crystallization}
that the quality of the random bond network of amorphous silicon is crucial to 
creating a high-quality sample.  Large amounts of hydrogen are used, not to 
saturate dangling bonds, but to reduce the number of strained bonds.  To date, however, 
it is unclear how this information about the deposition process relates to light-induced
degradation, primarily because it is unfeasible to isolate individual defects 
in the amorphous material for study.  Because of this, after 30 years of 
intense research aimed at understanding and mitigating this
relatively straightforward macroscopic effect, a complete {\em microscopic} explanation
for the performance degradation remains obscure. 

In this Letter, we investigate the effect of the silicon bond network on 
the formation of charge carrier traps.
Using a combination of classical force fields, density functional theory, and quantum Monte 
Carlo methods, we have sampled the space of random 
networks through bond switches, in which two silicon atoms exchange neighbors.
Our calculations demonstrate that bond switches can 
create dangling bonds and they also can create regions of strained bonds that trap 
holes.  These defects are preferentially formed when a region contains
both a hole and an excitation.  We show that a new picture of the 
SWE emerges from these results that is able to accomodate the characteristic
dependencies on temperature, pressure, and time observed in experiments.

We have used a set of methods that increase in accuracy and computational expense.
There is filtering done at each stage to ensure that the higher accuracy techniques 
are used only for potentially interesting samples of a-Si.  First the WWW\cite{www} process 
using the Keating\cite{keating} potential is 
performed, which filters the large space of all bond networks to a set of amorphous low-energy 
networks.  Each low-energy network is then perturbed by a single bond switch, and these
new networks are again filtered based on their energy.  These sets of bond networks 
are then passed to DFT using the Siesta\cite{siesta} program, where the geometries are 
optimized, which can potentially break bonds.  The hole trap depth is then obtained
by calculating the difference in ionization energies between the low-energy 
network and its perturbations, i.e., 
$( E_{perturbed}^+- E_{perturbed}^0)-(E_{reference}^+-E_{reference}^0)$.  Networks
with large trap depths are then singled out for analysis.  We use VASP\cite{vasp} with the
dimer method\cite{dimer_method} and nudged elastic band method\cite{neb_method} to find the transition state between the two bond networks 
for the ground and positively charged states.  Finally, diffusion Monte Carlo(DMC)\cite{Foulkes_review} using the QWalk\cite{qwalk}
program was used to evaluate
the transition barrier in the ground, positively charged, excited neutral, and 
excited positively charged states.
This approach has been carried out for both 64-atom and 216-atom periodically repeated cells.
  The distributions are in agreement for both sizes.
Because of the cost of finding many transition states and performing DMC calculations, 
we calculated only a few transition states in DFT on 216-atom cells, to confirm 
agreement with the 64-atom cells.  The DMC results were all on 64-atom cells.  All 
parameters were carefully checked for convergence\footnote{Energies were calculated
within SIESTA with a DZP basis, 300 meV energy shift, and PBE functional.  VASP 
calculations used a 245 eV energy cutoff, ultrasoft potentials, and PW-91 functional.
DMC calculations used a Slater-Jastrow trial function with the one-particle orbitals
from SIESTA with a TZTP basis, variance-minimized Jastrow coefficients, and  
a timestep of 0.05 Hartrees$^{-1}$}.

\begin{figure}
\includegraphics[width=\columnwidth]{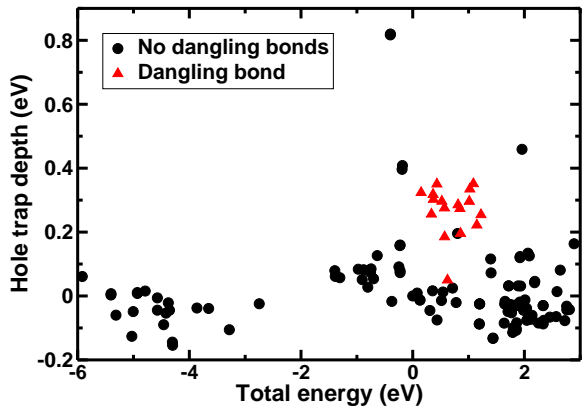}
\caption{Hole trap depths for 128 216-atom bond networks referenced to the network at (0,0),
 with dangling bond states (triangles) separated from 
systems with all four-fold coordinated Si atoms (circles).  While a dangling bond is usually a 
trap, there are strained bonds that trap hole very strongly.}
\label{table:traps}
\end{figure}

The statistical nature of this search is particularly crucial:
unlike a perfect crystal, in an amorphous material a single periodically repeated
simulation cell is not necessarily  representative of the entire 
phase space. Thus, we perform all possible bond switches for six independent a-Si samples
containing 216 atoms and twenty samples containing 64 atoms.  There 
are hundreds of possible switches in each sample, of which we keep those 
 that increase the energy less than 1.3 eV in our model.  
The results presented here thus are statistical, deriving from the analysis of 
hundreds of computational samples; we present representative examples to simplify 
the discussion, but the trends discussed throughout this work apply to the entire computational set.

After the classical model/DFT minimization procedure, we analyze the electronic states
to determine which network structures are hole traps/electron traps/low-energy absorbers 
(Fig ~\ref{table:traps}).
In our calculations, approximately 1-2 bond switches/nm$^3$ change the electronic levels
significantly from the reference structure, producing both electron and hole traps, as
well as absorption at lower energy than the nominal 1.8 eV gap of amorphous silicon.
We focus on the hole traps, since hole transport is the limiting factor in a-Si 
solar cells.

While the classical model used cannot break bonds, upon relaxation within 
DFT we find several instances where dangling bonds are formed, along with a
complementary silicon atom with five bonds (a floating bond).
A single bond switch is sufficient to form these dangling bond/floating bond pairs at either side
of the switch, which results in a separation of approximately 1 nm, and only slightly 
higher energy (around 0.1 eV) than the original structure.  This is similar in concept to the 
results of Biswas et al.\cite{biswas_network} using a simpler tight-binding treatment
of the electrons.

\begin{figure}
\includegraphics[width=\columnwidth]{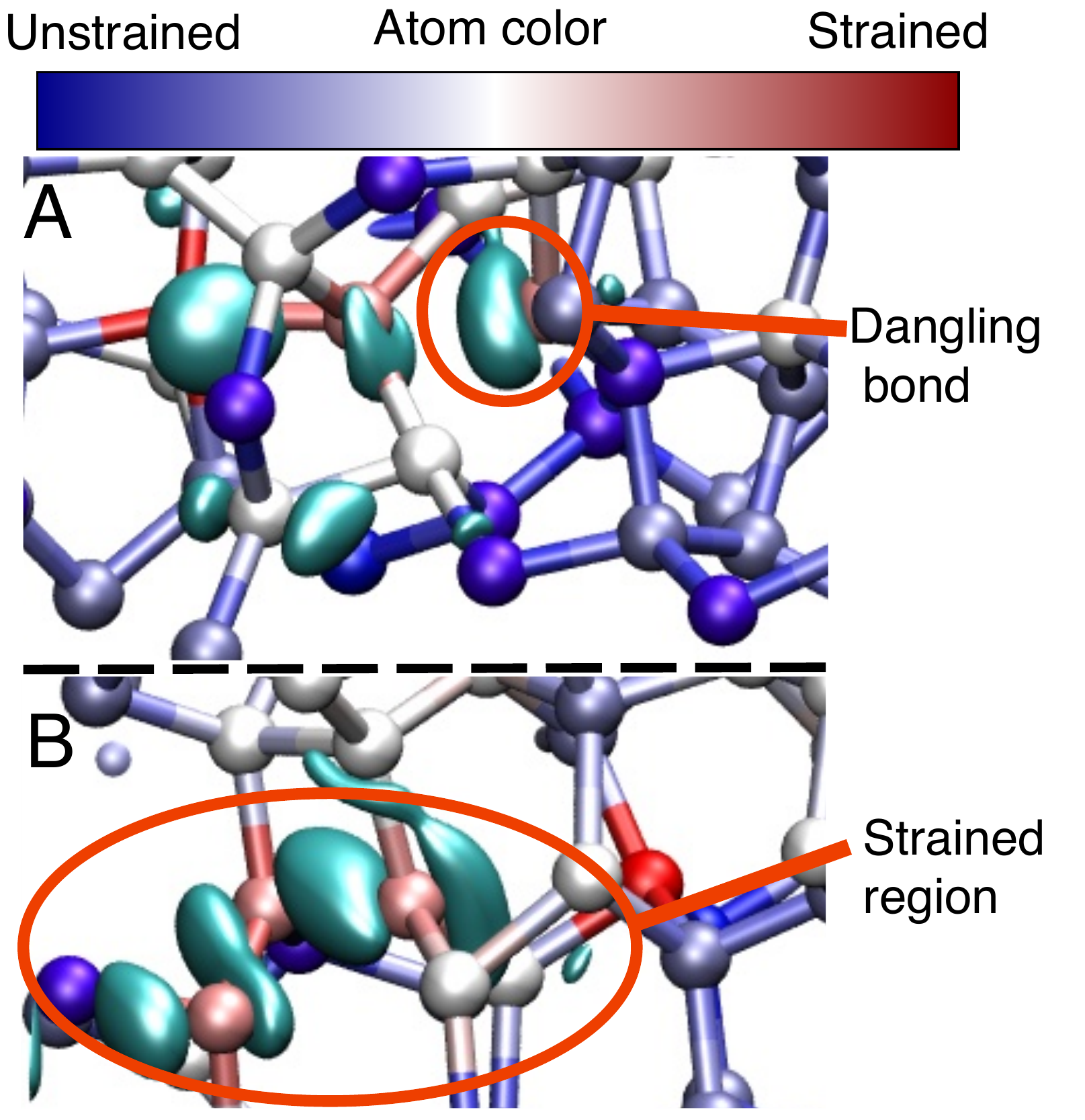}
\caption{Hole state in a-Si with a dangling bond (A) and 
without (B).  The amount of strain increases from blue to white to red.  
In both cases, the orbital (hole)
is centered around a region of strained silicon atoms (white to red).
Sample B (without dangling bonds) is lower in energy for a hole by 0.625 eV in our calculations.
The highly strained atom on the right side of sample B does not attract the hole state
because there are unstrained bonds surrounding it, which would cause the hole to be
in a highly confined, energetically unfavorable state. The atoms are colored according to 
their energy in the Tersoff\cite{tersoff} effective potential.}
\label{fig:homo_strained}
\end{figure}

Surprisingly, however, we find that the deepest traps are not dangling bonds.
 To help understand why, we investigate two representative samples shown in 
Figure ~\ref{fig:homo_strained}.   The presence of the dangling bond (Figure ~\ref{fig:homo_strained}A)
 allows nearby 
bonds to relax, since there are only three constraints on the 3-fold coordinated atom.
A hole would then be forced to localize on the one atom, which is energetically 
less favorable due to an increase in the kinetic energy.  The same reasoning holds true
 for a fully saturated but highly strained atom, as can be seen by the fact that the hole 
state does not localize on the highest strained atom in Figure ~\ref{fig:homo_strained}B.   On 
the other hand, 
a group of two to four strained atoms allow the hole to be less confined, while still 
binding a valence electron much more weakly (and thus binding a hole more strongly) than 
unstrained atoms.  The hole is localized, so this state is a hole trap.
 The region of strained bonds is a stronger trap because of quantum confinement of the hole.

These results establish that strained regions of silicon atoms could be as important
as dangling bonds for hole transport in amorphous silicon and that these defects can be created 
by a single bond switch.  We now turn to the mechanism in which these bond switches can occur, 
which requires detailed analysis of the reaction pathway.  
Reaction barriers are well-known to be poorly described by density functional theory; one must
employ a more accurate first principles method for a reliable description\cite{jeff_barriers}. 
 Thus, we use DFT to obtain the reaction path on 64-atom samples using the dimer
method\cite{dimer_method}, and then evaluate energy differences along
the path using the highly accurate fixed node DMC method. 
 The DMC barriers differ from DFT by up to 50\%, 
which is enough to change the picture significantly, so the more accurate calculation is 
necessary in this case.

\begin{figure}
\includegraphics[width=\columnwidth]{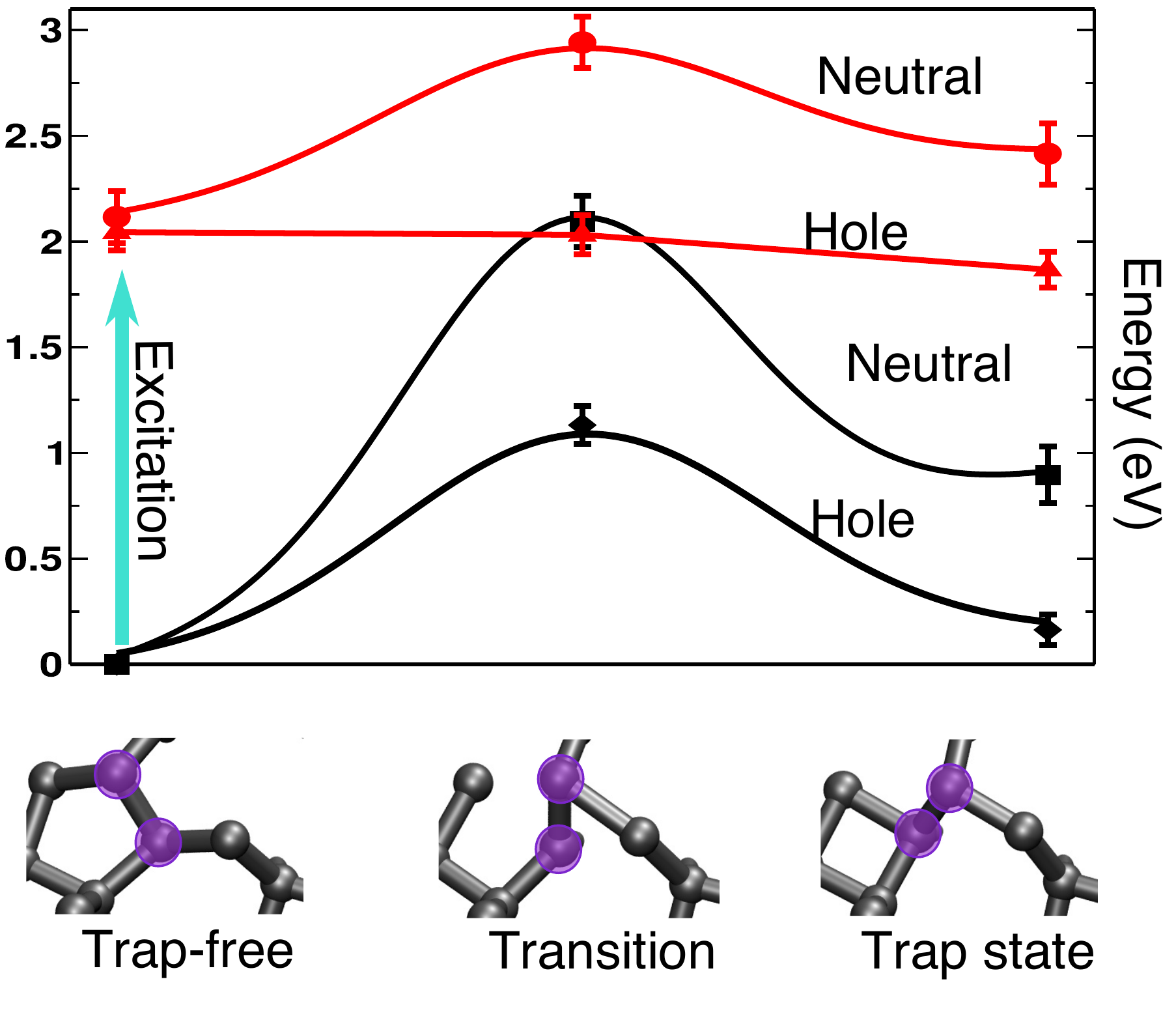}
\caption{Diffusion Monte Carlo energies as a sample of a-Si undergoes a bond 
switch process in both a ground and photo-excited state.  The highlighted atoms are exchanging their bonds.  The lines 
are guides to the eye.}
\label{fig:barrier}
\end{figure}

There are several potential light-induced events that may cause a bond switch.  One is 
 the collision and recombination of an electron and hole,
which effectively locally heats the bond network, enabling the large ground-state
 barrier to be overcome.  This mechanism is not supported by the high barrier and large 
energy cost in the 
ground state (Figure ~\ref{fig:barrier}).  Another possibility is through a light-induced
 electronic state other than the ground, neutral state.  For this possibility, there are three 
major states: a hole, an electron, and an excited state, any two of which can 
potentially exist simulaneously in a region.  Since holes are the slowest
charge carriers, taking around 250 ns to exit a 500 nm device\cite{pv_handbook}, 
they have the highest density of the three states.  The second most numerous state
is the excited state, which has a lifetime of around 10 ns\cite{exciton_lifetime}.
Finally, the electron exits a 500 nm device in around 1 ns\cite{pv_handbook}, so it is the most sparse.
The following analysis does not depend strongly on the actual lifetimes; only on their
ordering.

If a hole happens to be in the region and the product bond configuration 
is a hole trap compared to the initial configuration, the energy 
difference between configurations decreases by the difference in 
ionization energies, which can be around 0.1-0.7 eV according to our calculations.  
The presence of a hole therefore preferentially forms hole-attracting bond networks; 
however, the reaction barrier to forming such networks is still substantial (Figure ~\ref{fig:barrier}).  
One can imagine two further modifications to 
the electronic state--first, the system could attract a second hole, which is not 
favored electrostatically and in our calculations does not decrease the barrier 
very much, or second, the system 
could absorb a photon, exciting the unpaired electron.

When the unpaired electron is photoexcited, a single bond can lose one electron from the
hole and have the other excited to an anti-bonding state from the excitation, 
which allows a bond to break and switch.  In the case presented here in Figure ~\ref{fig:barrier}, which is 
not unique in our samples, the energy ordering changes and the barrier is reduced to zero.
Thus, when a hole is in a region that has the potential to change to 
a hole trap, and the region then absorbs a photon, a hole trap is formed
 with {\em no} barrier and a reduction in energy.  When the excitation dissociates or
decays, the energy ordering returns to the original condition and the barrier
reappears.

These calculations lead us to the following potentially important mechanism in the SWE.
A hole travels through a-Si 
slowly.  While it is in a region, bond network changes with hole traps 
become more energetically favorable, but there is still a large barrier for 
the bond switch necessary to change the bond network.
 If that region happens to absorb a photon, 
the barrier to switch bonds is zero or nearly so, and the network with a hole trap
is lower in energy.  The system then performs the bond switch, 
which leads to a hole trap state.

This explanation of the SWE has several implications about the macroscopic behavior of
a-Si, which can be validated by comparing to experiment.
We have shown that bond switches can form dangling bonds, so experiments that measure the
number of dangling bonds increasing with the decrease in solar cell efficiency are 
consistent with our results.  In our picture, however, the dangling bonds are 
{\em not} the only hole traps.  This is also in agreement with the observation that the 
number of dangling bonds can vary by a factor of ten and still have the same efficiency 
in the same sample\cite{shimizu_review}.  
Furthermore, since strained silicon bonds are the major cause, our picture
predicts a reduction in the SWE under a decrease in hydrostatic pressure, 
since the bonds are then able to relax.  Dangling bonds, on the other hand, 
are less affected by a change in pressure, since the lone electron is still present.  
 This provides a way of gauging the relative importance of the 
two types of defects. In experiments
on light soaking, a slight increase in 
the volume of the sample is observed\cite{volume_change}, which in our picture is due to an increase in strain.  
It has also been observed\cite{defect_strain,defect_relaxation}
 that if one removes the strain due to deposition, the SWE is reduced.

\begin{figure}
\includegraphics[width=\columnwidth]{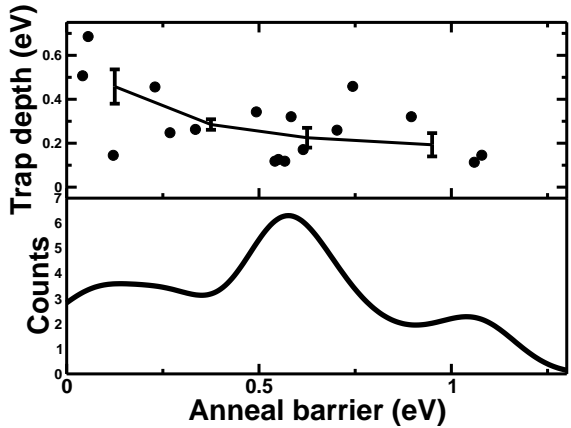}
\caption{(top) hole trap depth as a function of anneal barrier   The line is an average of the trap depth
for the surrounding region, with error bars indicating the rms deviations.
(bottom) the anneal barrier distribution
Numbers are from DFT with a uniform correction from DMC. }
\label{fig:anneal_barr}
\end{figure}

As an advantage of taking a statistical approach, we have the opportunity to 
compare the distribution of anneal barriers and trap depth.
On 19 hole traps, the depth of the trap is anticorrelated with the anneal barrier, 
and the distribution of anneal barriers has a wide peak at around 0.70 eV (Figure ~\ref{fig:anneal_barr}).
This anticorrelation is well-known in experiment; that is, traps formed at low temperatures
are deeper, and a very similar distribution of anneal energies explains the
temperature dependence of the SWE\cite{anneal_barriers}.

Based on our simulations, one can develop a simple rate model for the creation of traps.
Our results show that the barrier to create a hole trap 
 is zero when a hole and an excitation collide, so $\frac{dN}{dt} \propto xp$, 
where $N$ is the density of defective regions, $x$ is the concentration of excited states, 
and $p$ is the concentration of holes.
It is generally accepted that there is a concentration of holes proportional to $\frac{G}{N}$, 
where $G$ is the flux of photons.  Furthermore, 
the probability that a photon is absorbed by a region that is {\em not} defective is proportional
to $\frac{G}{N}$.  Therefore, the creation rate for a simultaneous hole/photon is
 $\frac{dN}{dt}\propto \frac{G}{N}\cdot \frac{G}{N}$, which has the solution 
$N \propto t^{1/3}G^{2/3}$ observed in experiment.
Since the barrier is often nearly zero with a hole and excitation present, this analysis 
suggests that the
defect creation rates in light would be nearly the same at all temperatures, which is the
 case in experiment\cite{fritzsche_temperature}, but difficult to explain using a model 
that depends soley on diffusion of defects.

This picture opens several avenues for mitigation of the SWE.  
One is to redesign deposition processes to reduce the internal stress in the thin film,
which reduces the number of hole traps.  In the same vein, nanoscale features in the 
material that allow stress to be relieved could largely mitigate the SWE.  
An opposite strategy is to increase the rigidity of the bond network to prevent 
bond switches from happening in the first place; for example, by embedding 
nanocrystals in the material\cite{nanocrystal_asi}. Finally, in our description of
the SWE the reaction is driven by an excitation combined with a hole that reverses
the energy ordering of the defect-free and defective states.  If a catalyst is
introduced that reduces the ground state barrier, the defective bond network 
that is higher energy in the ground state can relax into the defect-free network
at a lower than operating temperature, thus mitigating the SWE.

 We have found that holes are trapped by both strained silicon bonds and
dangling bonds and provided a simple and plausible mechanism in which the traps can form.
The depth of the hole traps is determined by a balance between quantum confinement of 
the hole and the amount of strain in the bond, for which accurate treatment of electrons
is particularly crucial.
The strained bond/bond switch explanation of the SWE suggests further exploration 
in experiments aimed at mitigating the SWE.  Advances in this area have the opportunity
of increasing the efficiency of amorphous silicon based solar cells by 30\% or more, which 
is a significant step towards creating a less expensive alternative to higher efficiency  crystalline 
silicon cells.  

\begin{acknowledgements}
This work was performed under the auspices of the National Science Foundation by the University 
of California Berkeley under Grant No. 0425914.
\end{acknowledgements}

\begin{footnotesize}

\bibliography{asi}
\end{footnotesize}

\end{document}